\documentclass[twocolumn,showpacs,preprintnumbers,amsmath,amssymb]{revtex4}

\usepackage{graphicx}
\usepackage{dcolumn}
\usepackage{bm}

\begin{document}

\newcommand{\be}{\begin{equation}}\newcommand{\ee}{\end{equation}}
\newcommand{\bear}{\begin{eqnarray}}\newcommand{\eear}{\end{eqnarray}} 
\newcommand{\ba}{\begin{array}}\newcommand{\ea}{\end{array}}
\newcommand{\lae}{\begin{array}{c}\,\sim\vspace{-1.55em}\\< \end{array}}
\newcommand{\gae}{\begin{array}{c}\,\sim\vspace{-1.55em}\\> \end{array}}

\newcommand{\ds}{\mbox{$D\hspace*{-.65em} / \; $}}

\preprint{FERMILAB-PUB-04-289-T}\preprint{hep-ph/0411004}

\title{Massless Gauge Bosons other than the Photon}
\author{Bogdan A.~Dobrescu}
\affiliation{Fermi National Accelerator Laboratory, Batavia, IL 60510, USA}

\date{November 1, 2004} 

  \begin{abstract}
Gauge bosons associated with unbroken gauge symmetries, under which 
all standard model fields are singlets, may interact 
with ordinary matter via higher-dimensional operators. 
A complete set of dimension-six operators involving a massless $U(1)$ 
field, $\gamma^\prime$, and standard model fields is presented. 
The $\mu \rightarrow e \gamma^\prime$ decay, primordial nucleosynthesis, 
star cooling and other phenomena set lower limits on the scale of 
chirality-flip operators in the 1 -- 15 TeV range, if
the operators have coefficients given by the corresponding Yukawa couplings.
Simple renormalizable models induce $\gamma^\prime$ interactions with leptons 
or quarks at two loops, and may provide a cold dark matter candidate. 
  \end{abstract}

\maketitle


Experiments have established the existence of 12 gauge bosons.
Although gauge invariance requires the gauge bosons to be massless,
three of these ($W^\pm,Z$) are massive due to spontaneous
symmetry breaking, while eight other (gluons) are confined in massive 
hadrons. The only massless spin-1 particle discovered so far is the photon.

This letter addresses the possibility that 
massless gauge bosons other than the photon exist. 
If any standard model field is charged under a new {\it unbroken} 
gauge symmetry, the requirements of fermion mass generation and gauge 
anomaly cancellation \cite{Appelquist:2002mw} force the new 
symmetry to be $U(1)_{B-L}$, and the nonobservation of long range 
forces other than electromagnetism and gravity imposes an extremely severe 
constraint on the new gauge coupling,
$g_{B-L} \ll m_n/ M_{\rm Pl} \approx 10^{-19}$,
where $m_n$ is the neutron mass, and $M_{\rm Pl}$ is the Planck scale
(neutrino screening cannot relax the bound \cite{Blinnikov:1995kp}).

It is possible, however, that standard model fields 
are neutral with respect to a gauge group, but nevertheless
interact with the new massless gauge bosons. 
This follows from the existence of gauge invariant operators, of mass-dimension 
six or higher, which involve gauge field strengths. Such interactions have a natural 
decoupling limit: rather than forcing a dimensionless parameter to be 
extremely small, the experimental limits translate into a lower bound on the 
mass scale that suppresses the higher-dimensional operators.

In the case of an Abelian gauge symmetry, $U(1)_p$,
under which all standard model fields are singlets,
there is a single renormalizable term in the Lagrangian that involves both
the $U(1)_p$ gauge boson and standard model fields: a kinetic mixing 
$c_0 P^{\mu\nu} B_{\mu\nu}$, where $B^{\mu\nu}$ is the $U(1)_Y$ (hypercharge) 
field strength, $P^{\mu\nu}$ is the $U(1)_p$ field strength, and $c_0$ is a 
dimensionless parameter.
One might worry that this kinetic mixing would induce
dimension-four couplings of the new gauge boson to all standard model  
fermions, but this is not the case \cite{Holdom:1985ag}: the kinetic terms of the 
two $U(1)$ fields can be diagonalized and canonically normalized 
by an $SL(2,R)$ transformation, and their ensuing global $SO(2)$ symmetry
allows the identification of the linear combination of $U(1)$ fields that couples 
to hypercharge as the new hypercharge gauge boson. 
The orthogonal combination, referred to 
as the ``paraphoton'' in Ref.~\cite{Holdom:1985ag} and denoted here 
by $\gamma^\prime$, does not
have any renormalizable coupling to standard model fields.

The leading interactions of the $U(1)_p$ field with standard model fields
are given by dimension-six operators, so they are suppressed by two powers of 
the mass scale $M$ where the operators are generated.
The terms in the Lagrangian of this type which involve fermions are given by
\bear
\label{Lagrangian}
\frac{1}{M^2} \,  P_{\mu\nu} && \hspace*{-1em} \left(
\overline{q}_{\! L} \sigma^{\mu\nu}  C_u \widetilde{H} u_R
+\overline{q}_{\! L} \sigma^{\mu\nu} C_d H d_R \right.
\nonumber \\ [0.5em]
&& + \,\left. \overline{l}_{\! L} \sigma^{\mu\nu} C_e  H e_R + {\rm h.c.} \right) ~.
\eear
The notation is as follows: $q_L, l_L$ are
quark and lepton doublets, $u_R, d_R$ are up- and down-type 
$SU(2)$-singlet quarks, $e_R$ are 
electrically-charged $SU(2)$-singlet leptons,
and $H$ is the Higgs doublet.
An index labeling the three fermion generations is implicit.
The $3\times 3$ matrices in flavor space,
$C_u, C_d, C_e$, have complex elements which 
are dimensionless parameters.

If ``right-handed neutrinos'' (gauge singlet fermions) are present, 
there is an additional dimension-six operator 
involving $P^{\mu\nu}$, as well as a lepton-number violating 
operator of dimension five:
$P_{\mu\nu} \overline{N^c}_R \sigma^{\mu\nu} N_R$.
Otherwise, the leading $\gamma^\prime$  interactions with neutrinos are
given by a dimension-seven operator, $P_{\mu\nu} H\overline{l^c}_{\! L}
\sigma^{\mu\nu} H l_{\! L}$.

Eq.~(\ref{Lagrangian})
displays a complete set of dimension-six operators  
involving $U(1)_p$ gauge fields and standard model fermions.
For example, operators
which involve the dual field strength,
$\widetilde{P}_{\mu\nu} \equiv \epsilon_{\mu\nu\lambda\tau} P^{\lambda\tau}$,
can be reduced to Eq.~(\ref{Lagrangian}) using the identities
\bear
\widetilde{P}_{\mu\nu} \sigma^{\mu\nu} 
& = & - 2 P_{\mu\nu} \sigma^{\mu\nu} \gamma_5 ~,
 \nonumber \\ [1em] 
\widetilde{P}_{\mu\nu} \gamma^\mu D^\nu
& = & P_{\mu\nu} \left(\sigma^{\mu\nu} \ds -  2i \gamma^\mu D^\nu \right) \gamma_5  ~,
\label{ident}
\eear
and field equations such as $i\ds q_L = \lambda_u \widetilde{H} u_R + \lambda_d H d_R$,
where $\lambda_{u,d}$ are Yukawa coupling matrices,
and the covariant derivatives $D^\nu$ contain only standard model 
gauge fields.
Furthermore, using field equations and integration by parts, 
and ignoring operators of dimension higher than six,
one can show that any chirality-preserving operator can be written in terms of 
operators (\ref{Lagrangian}). For example,
the following identity is valid up to a total derivative:
\be
iP_{\mu\nu}\overline{q}_{\! L} \gamma^{\mu} D^\nu q_L = 
\frac{-i}{4}P_{\mu\nu}\overline{q}_{\! L} \sigma^{\mu\nu} 
\left(\lambda_u \widetilde{H} u_R + \lambda_d H d_R \right) + {\rm h.c.}
\ee 

In addition to the interactions with quarks and leptons
shown in Eq.~(\ref{Lagrangian}), the $U(1)_p$ field
has purely bosonic interactions described by dimension-six operators:
\be
\label{bosonic}
\frac{1}{M^2} H^\dagger H \left( c_1 B_{\mu\nu} + \tilde{c}_1 \widetilde{B}_{\mu\nu} 
+ c_2 P_{\mu\nu} + \tilde{c}_2 \widetilde{P}_{\mu\nu} \right)  P^{\mu\nu}   ~.
\ee
These renormalize the $U(1)_Y \times U(1)_p$ gauge 
couplings, and include vertices with paraphotons,
Higgs bosons, photons, and $Z$ bosons. 
The dimensionless parameters $c_{1,2}$, $\tilde{c}_{1,2}$ are real. 
Other operators vanish
({\it e.g.},
$ P_{\mu\nu} D^\mu H^\dagger D^\nu H$ and $ P_{\mu\nu} G^\mu_\rho G^{\nu\rho}$,
where  $G^{\mu\nu}$ is the gluon field strength) or are total derivatives
({\it e.g.}, $ P_{\mu\nu} G^\mu_\rho \widetilde{G}^{\nu\rho}$).

The dimensionless coefficients of the dimension-six
operators can in principle have any value 
consistent with an effective theory description, namely below $\sim 4\pi$.
However, the operators (\ref{Lagrangian}) flip chirality, and although the 
origin of flavor structure in the standard model is not yet known, 
the elements of $C_u,C_d$ and $C_e$ 
are expected to be of the order of or smaller than the corresponding Yukawa couplings.

Eq.~(\ref{Lagrangian}) is written in the weak eigenstate basis, where the 
standard model Yukawa couplings are flavor nondiagonal.
In the mass eigenstate basis, obtained by acting with 
different unitary $3\times 3$ matrices
on the left- and right-handed fields,  
the chirality-flipping operators
change their flavor dependence:
$C_f \rightarrow C_f^\prime = U_L^f C_f U_R^{f \dagger}$, 
where $U_L^f$ and $U_R^f$ are the unitary matrices that diagonalize the 
masses of the $f=e,u,d$ fermions. Note that $U_L^{u\dagger} U_L^d$
is the CKM matrix, whereas the relation 
between the neutrino mixing angles and $U^e_L$ is looser.
The interactions of mass-eigenstate fermions, $f^\prime$, 
with the $U(1)_p$ field appear in the Lagrangian as follows: 
\be
\label{mass-basis}
\frac{v_h}{M^2} P_{\mu\nu}  \left[ \overline{f^\prime} \sigma^{\mu\nu} 
\left({\rm Re}C_f^\prime + i\, {\rm Im}C_f^\prime \, \gamma_5 \right) f^\prime \right] ~.
\ee
These terms proportional to the vacuum expectation value of 
the Higgs doublet, $v_h \approx 174$ GeV, represent 
magnetic- and electric-like dipole moment operators.


There are various phenomenological constraints on the $\gamma^\prime$ interactions.
First, the successful predictions of primordial nucleosynthesis limit the 
number $\Delta g_*$ of effective relativistic degrees of freedom contributed 
by new particles that are in thermal equilibrium at a temperature 
$T_{\rm BBN} \approx 1$ MeV. 
The maximum value of  $\Delta g_*$ is often expressed in terms 
of the maximum number of additional neutrino species allowed by the data:
$\Delta g_*^{\rm max} = (4/7)\Delta N_\nu^{\rm max} $.
The data on light element abundances exhibit some
inconsistencies which translate into an uncertainty on
$\Delta N_\nu^{\rm max}$. 
In Ref.~\cite{Cuoco:2003cu}, it is found that
$\Delta N_\nu^{\rm max} = 0.6$ at the $2\sigma$ level, 
while in Ref.~\cite{Barger:2003zg} the $2\sigma$ and $3\sigma$
contours in the $N_\nu$ versus baryon density
plane extend to $\Delta N_\nu^{\rm max} \approx 0.2$ and $0.5$,
respectively.
At any rate, the bound does not allow the two degrees of freedom
of a paraphoton in thermal equilibrium.
Thus, the paraphoton must decouple at $T_{\gamma^\prime} > T_{\rm BBN}$,
so that during nucleosynthesis it 
contributes an effective number of degrees of freedom \cite{Davidson:1991si}
\be
\Delta g_*(T_{\rm BBN}) = 2 \left[ \frac{g_*(T_{\rm BBN})}
{g_*(T_{\gamma^\prime})}\right]^{4/3} ~.
\ee
Given that $g_*(T_{\rm BBN}) = 43/4$, the lower bound on the number of 
relativistic degrees of freedom at freeze out is 
\be
g_*(T_{\gamma^\prime}) > 27.5 \times (\Delta N_\nu^{\rm max})^{-3/4} ~.
\ee

The number of degrees of freedom in the standard model is $g_* = 247/4$ just above
the temperature of the QCD phase transition, 
$T_{\rm QCD} = 150 - 180$ MeV \cite{Fodor:2001pe},
and $g_* = 69/4$ just below $T_{\rm QCD}$. Hence, $T_{\gamma^\prime} = 180$ MeV 
is allowed as long as $\Delta N_\nu^{\rm max} > 0.34$.
The freeze-out temperature would have to be higher by an order of magnitude, 
$T_{\gamma^\prime} > m_\tau$, if $\Delta N_\nu^{\rm max} \approx 0.25$.
Note that the bound on $\Delta N_\nu^{\rm max}$ from a fit to the 
cosmic microwave background does not apply to the paraphoton, because 
at the time of recombination $g_* = 2$ 
rendering the effective number of $\gamma^\prime$  degrees of freedom negligible 
[smaller by a factor of $(43/8)^{4/3}$ than at $T_{\rm BBN}$].

The interaction rate of the paraphoton with the thermal bath is given by 
\be
\Gamma (T) = \frac{2\zeta (3)}{\pi^2} T^3 \; \langle \sigma_{\gamma^\prime} \rangle ~,
\ee
where $\zeta (3) \approx 1.202$, and 
$\langle \sigma_{\gamma^\prime} \rangle$ is the thermally averaged cross section
for $\gamma^\prime$ interactions with the standard model particles that are 
in thermal equilibrium. 
At temperatures just above $T_{\rm QCD}$, 
the thermal bath includes light quarks ($u,d,s$), 
leptons ($e,\mu$), photons, gluons and neutrinos. The dominant $\gamma^\prime$
interactions  involve the heaviest fermions, namely $\mu$ and $s$. 
A detailed study of the $\gamma^\prime$ decoupling would entail
solving a set of coupled Boltzmann equations, but for the purpose of 
estimating the $\gamma^\prime$ couplings in terms of $T_{\gamma^\prime}$
it suffices to make a simple approximation: 
the $\gamma^\prime$ interaction rate, $\Gamma (T_{\gamma^\prime})$,
equals the expansion rate of the universe at freeze out,
\be
\Gamma (T_{\gamma^\prime}) \approx  \frac{T_{\gamma^\prime}^2}{M_{\rm Pl} }  \left(\frac{2\pi^3 }{45} 
g_*(T_{\gamma^\prime})\right)^{1/2} ~.
\ee

The $\gamma^\prime$  annihilation due to interactions with muons proceeds
through the processes shown in Fig.1. The parametric dependence
of the annihilation cross section is
\be
\langle \sigma_{\gamma^\prime} \rangle \sim \frac{\alpha c_\mu^2 m_\mu^2}{M^4} ~,
\ee
where $c_\mu = \left|(C_e^\prime)_{22}\right| v_h/m_\mu \lae O(1)$
sets the strength of the muon-paraphoton interaction,
and $\alpha$ is the fine structure constant.
Thus, the constraint on $T_{\gamma^\prime}$
results in a limit on the effective mass scale of the 
muon-paraphoton interaction,
\be
\frac{M}{\sqrt{c_\mu}} \approx 3.9 \; {\rm TeV} \times 
\left[g_*(T_{\gamma^\prime})\right]^{-1/8} \left(\frac{T_{\gamma^\prime}}{1 \; {\rm GeV} }\right)^{1/4} ~.
\label{effective}
\ee
The $T_{\gamma^\prime}^{1/4}$ dependence dampens the 
sensitivity of the limit to the approximations used here. 
Using $g_*(T_{\gamma^\prime}) = 255/4$, which includes the $\gamma^\prime$ 
degrees of freedom, and $T_{\gamma^\prime} \gae 180$ MeV,
Eq.~(\ref{effective}) gives $M/\sqrt{c_\mu}  \gae  1.5 \; {\rm TeV}$. 
The $\gamma^\prime \gamma \rightarrow s\bar{s}$,
$\gamma^\prime s \rightarrow \gamma s $ and 
$\gamma^\prime \bar{s} \rightarrow \gamma \bar{s}$ processes at $T_{\gamma^\prime}$ 
impose a limit stronger by a factor of approximately 
$(\sqrt{3}m_s/m_\mu)^{1/2} \approx 1.2$
on $M/\sqrt{c_s}$, where $c_s \equiv |(C_d^\prime)_{22}| v_h/m_s$.

\begin{figure}[t]
\begin{center}
\begin{picture}(100,150)(60,-100)
\thicklines
\multiput(0,0)(20, 0){2}{\qbezier(0,0)(5,5)(10,0)\qbezier(10,0)(15,-5)(20,0)}
\multiput(0,30)(20, 0){2}{\qbezier(0,0)(5,-5)(10,0)\qbezier(10,0)(15,5)(20,0)}
\put(40,0){\line(0, 1){30}}
\put(40,0){\line(1, 0){40}}
\put(40,30){\line(1, 0){40}}
\put(0,8){\small $\gamma$}
\put(0,38){\small $\gamma^\prime$}
\put(80,8){\small $\mu^-$}
\put(80,38){\small $\mu^+$}
\multiput(140,0)(20, 0){2}{\qbezier(0,0)(5,5)(10,0)\qbezier(10,0)(15,-5)(20,0)}
\multiput(140,30)(20, 0){2}{\qbezier(0,0)(5,-5)(10,0)\qbezier(10,0)(15,5)(20,0)}
\put(180,0){\line(0, 1){30}}
\put(180,0){\line(1, 0){40}}
\put(180,30){\line(1, 0){40}}
\put(140,8){\small $\gamma$}
\put(140,38){\small $\gamma^\prime$}
\put(220,8){\small $\mu^+$}
\put(220,38){\small $\mu^-$}
\end{picture}

\vspace{-7.7em}
\begin{picture}(100,150)(60,-100)
\multiput(0,10)(10, 10){2}{\qbezier(0,0)(0,4)(5,5)\qbezier(5,5)(9,5)(10,10)}
\put(20,30){\line(-1, 1){20}}
\put(20,30){\line(1, 0){40}}
\put(60,30){\line(1, 1){20}}
\multiput(60,30)(10, -10){2}{\qbezier(0,0)(0,-4)(5,-5)\qbezier(5,-5)(9,-5)(10,-10)}
\put(10,8){\small $\gamma^\prime$}
\put(11,48){\small $\mu^\pm$}
\put(68,8){\small $\gamma$}
\put(66,48){\small $\mu^\pm$}
\multiput(140,13)(20, 0){2}{\qbezier(0,0)(5,5)(10,0)\qbezier(10,0)(15,-5)(20,0)}
\put(140,43){\line(1, 0){40}}
\put(180,13){\line(0, 1){30}}
\put(180,13){\line(1, 0){40}}
\multiput(180,43)(20, 0){2}{\qbezier(0,0)(5,5)(10,0)\qbezier(10,0)(15,-5)(20,0)}
\put(140,21){\small $\gamma^\prime$}
\put(140,51){\small $\mu^\pm$}
\put(220,21){\small $\mu^\pm$}
\put(220,51){\small $\gamma$}
\end{picture}
\vspace*{-11em}
\caption{Paraphoton annihilation via muon pair production 
($\gamma^\prime\gamma \rightarrow \mu^+\mu^-$) and Compton-like processes
($\gamma^\prime\mu^\pm \rightarrow \gamma\mu^\pm$).
}
\end{center}
\vspace*{-2em}
\end{figure}
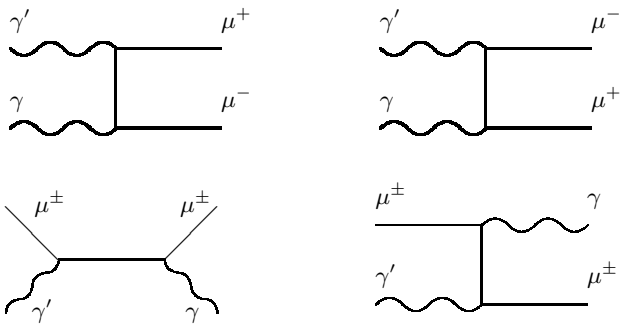


Another phenomenon that can be affected by paraphotons is star cooling.
One may use the studies
of axions in order to derive the energy loss in stars
due to $\gamma^\prime$  emission \cite{Hoffmann:1987et}.
The axion is similar to the paraphoton: they are 
bosons with derivative couplings to fermions.
The axion though has spin zero rather than 
one, so that the energy loss from stars is twice larger for 
$\gamma^\prime$ emission than for axion emission 
when the effective couplings are equal \cite{Raffelt:wa}.
The energy loss due to electron-paraphoton interactions 
is proportional to the square of $g_{e\gamma^\prime} = 4c_e m_e^2/M^2$, 
where $c_e \equiv |(C_e^\prime)_{11}| v_h/m_e$.
The limit on $\gamma^\prime$ emission through Bremsstrahlung, such as 
$e^- + ^{4}\!\hspace*{-0.05em}{\rm He} \rightarrow e^- 
+ ^{4}\!\hspace*{-0.05em}{\rm He} + \gamma^\prime$,
from the core of red giant stars \cite{Raffelt:wa} requires
$g_{e\gamma^\prime}^2/(4\pi) < 2.5\times 10^{-27}$, 
 so that $M/\sqrt{c_e} \gae 3.2$ TeV.
Compton-like scattering, $\gamma e^- \rightarrow \gamma^\prime e^-$,  
in horizontal-branch stars sets a slightly weaker limit,
$M/\sqrt{c_e} \gae 1.8$ TeV.

The neutrino signal from supernova 1987A
also limits the cooling through $\gamma^\prime$ emission,
setting a bound on the coupling of 
the paraphoton to nucleons. The leading 
nucleon-paraphoton interactions are similar to Eq.~(\ref{mass-basis}):
\be
\label{nucleon}
\frac{v_h}{M^2} P_{\mu\nu}  \overline{\cal N} \sigma^{\mu\nu} 
\left({\rm Re}C_{\cal N} + i {\rm Im}C_{\cal N} \gamma_5 \right) {\cal N} ~.
\ee
The $C_{\cal N}$ form factors are of the order of $(C_d^\prime)_{11}$ or 
$(C_u^\prime)_{11}$. For example, 
QCD sum rules give ${\rm Im}C_{\cal N} \approx 0.2 \, {\rm Im}(4 C_d^\prime - C_u^\prime)_{11}$
for the neutron \cite{Pospelov:2000bw}. The effective nucleon-paraphoton 
coupling, $g_{{\cal N}\gamma^\prime}$, is proportional to the nucleon mass:
$g_{{\cal N}\gamma^\prime} \approx 4c_n m_d m_n/M^2$, where 
$c_n \equiv |C_{\cal N}|v_h/m_d \lae O(1)$. Requiring that the 
supernova was cooled predominantly by neutrinos implies
$g_{{\cal N}\gamma^\prime} \lae 2\times 10^{-10}$, 
so that  $M/\sqrt{c_n} \gae 7$ TeV.
There is also a range of larger nucleon-paraphoton couplings 
($2\times 10^{-7} \lae g_{{\cal N}\gamma^\prime}  \lae 7 \times 10^{-5}$)
allowed by the supernova 1987A signal: the 
paraphotons are there too strongly coupled to nucleons 
to escape easily the supernova, and too weakly coupled 
to produce a signal in the detectors that 
measured the neutrino signal. However, that range
is not compatible with the nucleosynthesis constraint on
quark-paraphoton couplings.


The long-range forces induced by paraphoton exchange 
between chunks of ordinary matter are feeble. 
The interactions of nonrelativistic electrons or nucleons with 
the paraphoton are spin dependent.
Given that the average spin of macroscopic objects is 
very small, the lower limits on $M/\sqrt{c_e}$ or $M/\sqrt{c_n}$ imposed 
by measurements of long range forces are evidently 
weaker than the TeV scale. 

The magnetic- and electric-like dipole moments of the $t$ quark
could be probed at the Tevatron, LHC or a linear $e^+e^-$
collider if the dimensionless coefficient $|(C_u)_{33}|$ is of order unity and 
the mass scale $M$ is of order 1 TeV.


The flavor off-diagonal interactions included in Eq.~(\ref{mass-basis})
are constrained by various flavor changing neutral current processes.
The most severe limit comes from 
the $\mu \rightarrow e \gamma^\prime$ decay, and 
depends on $c_{e\mu} \equiv |(C_e^\prime)_{12}| v_h/m_\mu$. 
A comparison of the decay width, 
\be
\Gamma \left( \mu \rightarrow e \gamma^\prime \right) 
= c_{e\mu}^2 \frac{m_\mu^5}{8\pi M^4}  ~,
\ee
times the measured muon lifetime, $3.3\times 10^{9}$ eV$^{-1}$, 
to the experimental limit shown in Fig.~2 of Ref.~\cite{Bryman:1986wn},
${\rm Br}\left( \mu^+ \rightarrow e^+ X \right) 
< 3 \times 10^{-5}$ for any massless particle $X$, leads to
\be
\label{cemu}
\frac{M}{\sqrt{c_{e\mu}}} \gae 15 \; {\rm TeV} ~.
\ee
Flavor-changing neutral current processes in the hadron sector are 
also affected by $\gamma^\prime$ emission. Note though that the
$K^+ \rightarrow \pi^+ \gamma^\prime$ decay is forbidden by angular momentum 
conservation, while the experimental constraints on other processes such as 
$\Sigma^+ \rightarrow p \gamma^\prime$ or $B \rightarrow K^*\gamma^\prime$,
where $\gamma^\prime$ carries missing energy, are quite weak. 

In the presence of kinetic mixing of the $U(1)_p \times U(1)_Y$ fields,
the operators (\ref{Lagrangian}) contribute to the magnetic and 
electric dipole moments of the quarks and leptons.
Hence, there are constraints on the
products of the kinetic mixing parameter $c_0$ and the coefficients
of operators (\ref{Lagrangian}). 
For example, the magnetic moment of the muon is shifted by 
$\sim 4 c_0 c_\mu (m_\mu / M)^2/e$,
and requiring that this is less than $10^{-9}$ gives
$M/\sqrt{c_0 c_\mu } \gae 12$ TeV. 


The bosonic operators (\ref{bosonic}) lead to non-standard Higgs boson decays, 
$h\rightarrow \gamma\gamma^\prime$, $h\rightarrow Z\gamma^\prime$ and  
$h\rightarrow \gamma^\prime\gamma^\prime$, with $\gamma^\prime$ behaving in 
detectors as missing transverse energy. 
If the mass scale $M$ is not much higher than the electroweak scale
and the parameters $c_1$ and $\tilde{c}_1$ are not much less than unity,
a decay $h\rightarrow \gamma \,\slash{\!\!\!\! E_T}$ could produce 
striking signals in collider experiments. 

In the case of a non-Abelian gauge symmetry with respect to which 
all standard model fields are singlets, gauge invariance forces the operators 
describing interactions of the new gauge bosons to 
include two field strengths. As a result, the new gauge bosons
can be produced only in groups of two or more, 
and most operators have  mass dimension eight
or higher. The only exceptions are the operators analogous to 
the last two terms in Eq.~(\ref{bosonic}),
which lead to invisible Higgs decays.


The operators shown in Eq.~(\ref{Lagrangian}) describe
physics only at scales below $\sim M$.
However, simple renormalizable models generate these
operators after integrating out some heavy states.
A generic feature of these models is the presence of fields charged 
under $U(1)_p$. The lightest particle of this type is stable and 
may be a viable dark matter candidate \cite{Davidson:1991si}
provided it does not carry color or electric charge. It turns out though
that even if that particle is a singlet under $U(1)_Y$, 
its electric charge cannot be exactly zero.
This is because a kinetic mixing of the $U(1)_p\times U(1)_Y$ gauge bosons 
is induced at some loop level, and the $SL(2,R)$ and $SO(2)$ 
transformations that diagonalize the kinetic terms shift the electric 
charges of all fields carrying $U(1)_p$ charge \cite{Holdom:1985ag}.
A dark matter particle with mass of order 1 TeV must have an electric charge 
less than $10^{-4}$, because otherwise it would have left 
an imprint on the cosmic microwave background \cite{Dubovsky:2003yn}.

In models where the dimension-six operators arise at one loop, there is always a 
field charged under both $U(1)_p$ and $U(1)_Y$. A loop with this field 
in the internal line would induce a gauge kinetic mixing 
which in turn would lead to an electric charge for the 
dark matter particle that may be larger than $10^{-4}$.
This problem is avoided in some models that generate the dimension-six operators 
at two loops: the loop-induced kinetic mixing is negligible if all
fields charged under $U(1)_p$ have zero hypercharge, and the renormalizable
tree-level kinetic mixing is absent if one of the 
$U(1)$'s is embedded in a non-Abelian group.

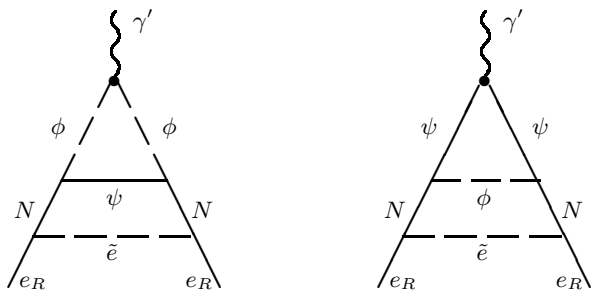
\begin{figure}[t!]
\begin{center}
\begin{picture}(200,150)(70,-40)
\thicklines
\put(80,40){\line(1, 0){40}}
\put(80,40){\line(-1, -2){20}}
\put(120,40){\line(1, -2){20}}
\multiput(80,40)(7, 14){3}{\line(1, 2){5}}
\multiput(120,40)(-7, 14){3}{\line(-1, 2){5}}
\multiput(69,19)(16, 0){4}{\line(1, 0){12}}
\put(100,78){\circle*{4}}
\put(97,31){\small $\psi$}
\put(62,26){\small $N$}
\put(129,26){\small $N$}
\put(76,58){$\phi$}
\put(118,58){$\phi$}
\put(97,10){$\tilde{e}$}
\put(64,0){$e_{R}$}
\put(127,0){$e_{R}$}
\put(107,98){$\gamma^\prime$}
\multiput(100,80)(0,12){2}{\qbezier(0,0)(3,3)(0,6)\qbezier(0,6)(-3,9)(0,12)}
\multiput(220,40)(15, 0){3}{\line(1, 0){11}}
\put(220,40){\line(-1, -2){20}}
\put(260,40){\line(1, -2){20}}
\put(220,40){\line(1, 2){18}}
\put(260,40){\line(-1, 2){18}}
\multiput(209,19)(16, 0){4}{\line(1, 0){12}}
\put(240,78){\circle*{4}}
\put(237,31){\small $\phi$}
\put(202,26){\small $N$}
\put(269,26){\small $N$}
\put(216,58){$\psi$}
\put(258,58){$\psi$}
\put(237,10){$\tilde{e}$}
\put(204,0){$e_{R}$}
\put(267,0){$e_{R}$}
\put(247,98){$\gamma^\prime$}
\multiput(240,80)(0,12){2}{\qbezier(0,0)(3,3)(0,6)\qbezier(0,6)(-3,9)(0,12)}
\end{picture}
\vspace*{-3.6em}
\caption{Electron-paraphoton interaction induced in a renormalizable model.}
\end{center}
\vspace*{-2.2em}
\end{figure}
As an example, consider a new scalar, $\tilde{e}$, with the same gauge charges 
as $e_R$, a gauge singlet Dirac fermion $N$, and a $\tilde{e}\,\overline{e}_R N$ 
coupling.
If in addition there is a scalar $\phi$ and a fermion $\psi$ charged only under
$U(1)_p$, and a $\phi\overline{\psi}N$ coupling, then 
$e_R$ couples to the paraphoton at two loops, as shown in Fig.~2. 
A Higgs Yukawa coupling inserted on an $e_R$ external line generates 
the last operator in Eq.~(\ref{Lagrangian}). 
If the new particles have masses of order $M$, and 
the new gauge and Yukawa couplings are of order unity, 
then $c_e, c_\mu, c_{e\mu}$ are given by a two-loop factor of order $(4\pi)^{-4}$.
Consequently, the limit on $M$ is rather loose: $M\gae 100$ GeV
from Eq.~(\ref{cemu}). 
The only  $U(1)_p$-charged particles are $\phi$ and $\psi$, 
and the lightest of them is a cold dark matter candidate.
Further studies are necessary to determine the region of parameter space
where the dark matter halo does not collapse too fast due to $\gamma^\prime$ 
emission.

The quark-paraphoton 
moments are induced in this model only at three loops and are 
negligible. Similar models, with $\tilde{e}$ replaced by 
a scalar having the same charges as one of the quark fields, 
induce at two loops only the quark-paraphoton 
moments, and the tightest limit on $M$ in that
case is set by $\gamma^\prime$ emission from supernova 1987A:
$M\gae 40$ GeV for $c_n \sim (4\pi)^{-4}$.
In the presence of an $H^\dagger H \phi^\dagger \phi$ coupling, 
the last two operators in Eq.~(\ref{bosonic}) are induced at one loop, and the 
gauge kinetic mixing arises at two loops.

It is intriguing that  massless gauge bosons other 
than the photon may interact with ordinary matter. 
The rather weak bound, below the electroweak scale in perturbative models, 
on the scale $M$ that suppresses such interactions 
makes it possible to search in collider experiments for the underlying 
dynamics that generate the dimension-six operators.
It would also be interesting to investigate alternative experimental
methods of searching for massless gauge bosons.

\medskip

{} {\it Acknowledgments:} \ 
I am grateful to William Bardeen, John Beacom, Gianfranco Bertone,
Sekhar Chivukula and Irina Mocioiu for helpful comments. 
This work was supported by DOE under contract DE-FG02-92ER-40704. 

\end{document}